# Polynomial time estimates for #SAT

Bernd R. Schuh

Dr. Bernd Schuh, D-50968 Köln, Germany; bernd.schuh@netcologne.de

**Abstract.**

Limits on the number of satisfying assignments for CNS instances with $n$ variables and $m$ clauses are derived from various inequalities. Some bounds can be calculated in polynomial time, sharper bounds demand information about the distribution of the number of unsatisfied clauses, $u$. Quite generally, the number of satisfying assignments turns out to be limited by $2^n \sigma_u^2 / (E(u)^2 + \sigma_u^2)$ where $\sigma_u^2$ is the variance and $E(u)$ the mean of this distribution. For large formulae, m>>1, bounds vary with $2^n/n$, so they may be of use only for instances with a large number of satisfying assignments.

## I. Introduction and notation.

SAT instances with $n$ atoms may have $O(2^n)$ many satisfying assignments, but may as well have only 1 or even 0 satisfying assignments or "solutions". In other words, the number of SAT solutions spans many orders of magnitude from 0 to $2^n$. In this sense satisfiability is decoupled completely from the size of the instance. That is why the decision problem $F \in$ SAT ? and the determination of specific solutions of $F$, #SAT, are problems of different complexity, namely NP and #P.

In the following I will formulate restrictions on #SAT by means of inequalities which can be evaluated in polynomial time (p.t.). Such restrictions only give rough limits on the number of solutions of a given SAT instance. Refinements are discussed for which p.t. performance is lost, however.

The central concept is the number of unsatisfied assignments $u(x)$, considered as a random variable over the set of equally probable assignments $x$ of a given SAT instance $F$. I will first set up notations in this section, then discuss the simplest bound in section II. In section III this bound will be sharpened by means of an adjustable parameter. Sections IV and V discuss choices for this parameter. Finally the limit of large instances is considered in section VI.

Except for tautologies propositional formulae can be written as a conjunction of clauses, i.e. in conjunctive normal form, CNF. Clauses are disjunctions of $n$ basic variables (atoms $a_s$) and their negations $(\overline{a_s})$. Assigning truth values (true/false=1/0) to each of the atoms defines an overall assignment $T_x$ to an instance $F$ which determines the satisfiability of $F$ under this particular assignment. The fraction of assignments which satisfy all clauses (and thus $F$) can be calculated via:

(I. 1) $\quad \nu_0 = 2^{-n} \sum_x T_x(F)$

A clause which is not satisfied by an assignment will be called "frustrated". Quite generally, by $\nu_i$ we denote the fraction of assignments which frustrate exactly $i$ clauses, $i \in \{0, 1, ..., m\}$.

(I. 2) $\quad \nu_i = 2^{-n} \sum_x \delta_{u(x),i}$

The Kronecker-$\delta$ picks out those assignments, for which $u(x)=i$. Thus $u(x)$ is the number of clauses $C_j$ frustrated by assignment $T_x$. This number can be calculated explicitly from:

(I. 3) $\quad u(x) = \sum_{j=1}^m T_x(\overline{C_j}) = \sum_{j=1}^m 2^{-k_j} \prod_{s=1}^n (1 - f_{js} x_s)$

Here, the $2^n$ truth assignments $T$ are numbered by $n$-tupels $x = (x_1, x_2, ..., x_n)$, where $x_j = 2T(a_j)-1 \in \{1, -1\}$. $k_j$ is the number of non-zero literals in clause j, and the adjacency-matrix element $f_{js}$ is +1 or -1 if atom $a_s$ or its negation $\overline{a_s}$ appears in clause j, and 0 otherwise. I will call the $x$ "states" in the following, and "solutions" (to $F$) if $T_x(F)=1$.

For 3-SAT (i.e. all $k_j<4$) one can calculate the quantity on the right hand side explicitly from

(I. 4a) $\quad u(x) = C - \sum_{s=1}^n \lambda_s x_s + \sum_{<st>} \mu_{st} x_s x_t - \sum_{<rst>} \nu_{rst} x_r x_s x_t$

where the coefficients are calculated from the adjacency matrix $f_{is}$ via

(I. 4b) $\quad C = \sum_{l=1}^m 2^{-k_l} \qquad \lambda_s = \sum_{l=1}^m 2^{-k_l} f_{ls} \qquad \mu_{st} = \sum_{l=1}^m 2^{-k_l} f_{ls} f_{lt} \qquad \nu_{str} = \sum_{l=1}^m 2^{-k_l} f_{lr} f_{ls} f_{lt}$

The brackets <> indicate that no element in the summation appears more than once. For a derivation see [1]. Obviously all these calculations can be performed in p.t..

## II. Basic bound.

Consider $u$ as a random variable on the space of all states $x$. Note that $u$ can take integer values only, including 0. Then (see (I. 2))

(II.1)    $\nu_i = prob(u = i)$  and  $\sum_{i=0}^{u_{max}} \nu_i = 1$ ,

and the expectation value $E(...)$  of any function $f$ of the variable $u$ can be written

$$E(f(u)) = 2^{-n} \sum_x f(u(x)) = \sum_{k=0}^{u_{max}} \nu_k f(k) \ .$$

The number of $x$ which fulfill $u=0$ equals the number of solutions to $F$, see (I. 1). Therefore

(II. 2)    $F \in$ SAT  iff   $\nu_0 \geq 1/2^n$ .

One can show easily that

(II. 3)    $\nu_0 \leq 1 - E(u)^2 / E(u^2) = \dfrac{\sigma_u^2}{\sigma_u^2 + E(u)^2}$  .

where  $\sigma_u^2$  is the variance of $u$.

There are various ways to derive this inequality. The Cauchy-Schwarz-inequality for random variables $A$ and $B$

(II. 4)    $E(AB)^2 \leq E(A^2)E(B^2)$

applied to $A=u$ and the indicator function $B$ defined by $B=1$ for $u>0$ and $B=0$ for $u=0$ yields:

$$E(u)^2 \leq E(u^2)(2^{-n} \sum_{x/u>0} 1) = E(u^2) \sum_{i>0} \nu_i = E(u^2)(1 - \nu_0) \ .$$

which is equivalent to (II. 3).  Another approach makes use of the formula

(II.5)    $2E(AC)E(BC) \leq E(C^2)\left( E(AB) + \sqrt{E(A^2)E(B^2)} \right)$

for three random variables $A$, $B$, $C$.  [3]

Set $C=0$ iff $u(x)=k$, 1 otherwise and define $A=uC=B$ . Then the inequality above leads to

(II. 6)    $\nu_k \leq \dfrac{\sigma_u^2}{\sigma_u^2 + (E(u) - k)^2}$

quite generally. (II. 3) is a special case. This derivation hints on the fact, that (II.3) is not a very sharp condition, because Cantelli's inequality

(II. 7)    $prob(X \geq E(X) + \lambda) \leq \dfrac{\sigma_X^2}{\sigma_X^2 + \lambda^2}$

for a random variable $X$ with variance  $\sigma_X^2$ and expectation $E(X)$ and a positive $\lambda$ tells us that

(II. 8)    $\nu_k \leq \nu_k + \nu_{k+1} + ... = prob(u \geq k) = prob(u \geq E(u) + (k - E(u))) \leq \dfrac{\sigma_u^2}{\sigma_u^2 + (E(u) - k)^2}$ .

Thus (II. 6) not only holds for a single $\nu_k$ but for the complete tail of the distribution, provided $k$ is larger than $E(u)$.

Quite generally, see equ. (I. 4a), $u(x)$ is a sum of products of the $x_i$. Since $x_i \in \{-1, +1\}$ the expectation value of odd products of the $x_j$ vanishes. Thus $E(u) = C$, as defined in (I. 4). For 3-SAT the variance can be expressed simply as:

$$\text{(II.9)} \quad \sigma_u^2 = \sum_s \lambda_s^2 + \sum_{\langle s,t \rangle} \mu_{st}^2 + \sum_{\langle r,s,t \rangle} \nu_{rst}^2 \ .$$

Again, all quantities so far can be calculated in p.t. .

To illustrate the procedure take a simple test instance

$$F \triangleq \begin{matrix} + & - & 0 \\ - & - & - \\ 0 & + & + \end{matrix}$$

which is the adjacency form of $F = (a_1 \vee \overline{a_2}) \wedge (\overline{a_1} \vee \overline{a_2} \vee \overline{a_3}) \wedge (a_2 \vee a_3)$ . Symbols + and − have been used instead of 1 and -1. The quantities in (I. 4) are easily calculated

$$\lambda_s = (1/8, -1/8, 1/8) \ ; \quad \mu_{st} = \begin{pmatrix} 0 & -1/8 & 1/8 \\ -1/8 & 0 & 3/8 \\ 1/8 & 3/8 & 0 \end{pmatrix}; \quad \nu_{123} = -1/8; \ all \ other \ \nu_{rst} = 0$$

Consequently $\sigma_u^2 = 15/64$; $C = 5/8$ and thus $\nu_0 \leq 3/8$. This is a very good estimate, since $F$ has exactly three solutions, i.e. $\nu_0 = 3/8$. This result is not surprising since the maximum value of $u$ is 1, thus $\overline{u^2} = \overline{u} = 1 - \nu_0$ and the upper limit equals $\nu_0$ trivially.

A somewhat more complicated instance with 4 atoms and 8 clauses is $F_{ex}$ given by

$$\begin{matrix} - & 0 & + & - \\ + & 0 & - & + \\ + & + & 0 & 0 \\ - & - & 0 & 0 \\ 0 & - & - & 0 \\ 0 & - & + & - \\ - & + & 0 & - \\ 0 & + & 0 & + \end{matrix}$$

For $F_{ex}$ one calculates $\sigma_u^2 = 5/8$ and $C = 3/2$ from the adjacency matrix , and (II. 3) yields

$$\nu_0 \leq 5/23 = 0.217..$$

Multiplied by $2^n = 16$ this gives 3.478.. as an upper bound for the number of solutions, which is a 300%-overestimate, since the actual number of solutions is 1 (0100 in an obvious notation). The next section will show, how the estimate can be improved.

**III. Sharpened bound.**

For an improvement of the inequality (II. 3) we use the sharpened Cauchy-Schwarz-inequality (sCS) [2]:

(III. 1)    $E(AB)^2 \leq E(\min(A^2, B^2)) E(\max(A^2, B^2))$

and apply it to $A = u / a$ and the indicator function $B$ as before. The "cutoff" $a > 0$ is introduced as a free parameter to optimize estimates.

To simplify notation let us introduce the abbreviations $\mu \equiv E(u) / a$ and $\beta \equiv E(u^2) / E(u)^2$ in the following. Then the basic inequality (II. 3) reads $v_0 \leq 1 - 1 / \beta$, whereas sCS leads to

(III. 2)    $v_0 \leq 1 - \Delta_\leq - \dfrac{\mu^2}{\beta\mu^2 + \Delta_\leq} = 1 - \dfrac{1}{\beta} - \Delta_\leq (1 - \dfrac{1/\beta}{\beta\mu^2 + \Delta_\leq})$ .

Here $\Delta_\leq = \Delta_\leq (a)$ is given by

(III. 3)    $\Delta_\leq = 2^{-n} \sum\limits_{x \in S_\leq} (1 - \dfrac{u(x)^2}{a^2}) = \sum\limits_{k=1}^{\lfloor a \rfloor} v_k (1 - \dfrac{k^2}{a^2})$   .

The summation extends over the set $S_\leq$ of all states $x$ for which $0 < u(x) \leq a$. $\lfloor a \rfloor$ denotes the largest integer smaller than $a$. Whether (III. 2) is a better estimate than (II.3) depends on the sign of the factor $(1 - \dfrac{1/\beta}{\beta\mu^2 + \Delta_\leq})$. This factor is positive for all choices of $a$. The proof is given in Appendix B.

Equation (III. 2) involves two intrinsic parameters, which are specific of the instance under investigation, namely the first two moments $E(u^2)$ and $E(u)$. Both can be calculated in p.t.. In addition there is the adjustable cutoff $a$ which can be used to optimize the estimate. On the other hand this bound also requires the knowledge of the distribution $\{v_k\}$ to calculate $\Delta_\leq$. Thus polynomial time calculability in general is lost.

A way out is to search for an approximation for $\Delta_\leq$ which can be calculated in p.t and with certainty gives a better estimate to $v_0$ as (II. 3). Since the right hand side of (III. 2), considered as a function of $x = \Delta_\leq$ is a monotonically decreasing function of $x$ for $x > x_{\max} = \mu(1 - \beta\mu)$ for all $a$ , (III. 2) remains to be an improvement on (II. 3) if one replaces the actual $\Delta_\leq$ by any approximative $\Delta_{app}$ which fulfills

(III. 4)    $\Delta_\leq \geq \Delta_{app} > \dfrac{1}{\beta} - \beta\mu^2$  .

We have also tried Stephen Walkers recently published "self-improved" CS-inequality [4], but found no improvement in our case. [5]

## IV. Medium cutoff

As an instructive example of the improvement via sCS we fix the cutoff $a$ at a value beyond the mean $E(u)$ of the distribution, namely

(IV.1)   $a = E(u) + \dfrac{\sigma_u^{\;2}}{E(u)}$

corresponding to $\beta\mu = 1$ . As a result we will find a relation between the number of solutions and a part (tail) of the distribution. With (IV. 1) equation (III.2) contracts to

(IV.2)   $\nu_0 \leq g(\beta\Delta_\leq)$   $with$   $g(x) = 1 - \dfrac{1}{\beta}(1 + \dfrac{x^2}{1+x})$ .

$g$ is a monotonically decreasing function, with $g(0)$= 1-1/β. Thus any approximation $0 < \Delta_{app} \leq \Delta_\leq$ will lead to $\nu_0 \leq g(\beta\Delta_\leq) \leq g(\beta\Delta_{app})$ .

We can approximate   $a^2\Delta_\leq(a) = \sum_{k=1}^{M}\nu_k(a^2 - k^2) = a^2\,\nu_\leq - \sum_{k=1}^{M}\nu_k k^2 \geq (a^2 - M^2)\nu_\leq$   with $M$ denoting the largest integer below $E(u^2)/E(u)$ .

We define

(IV.3)   $x_{app} = \dfrac{a^2 - M^2}{E(u^2)}\nu_\leq = [\beta - M^2/E(u^2)]\nu_\leq$ ;   with $\nu_\leq = \nu_1 + \nu_2 + ... + \nu_M$

Then   $\nu_0 \leq g(x_{app})$   relates the number of solutions to $\nu_\leq$, i.e. the probability of $u$ lying below $a$.

Checking the estimate with this approximation for $F_{ex}$ yields $M$=1 ($a$=23/12), thus $\nu_\leq = \nu_1$ =1/2, $x_{app} = 385/828 = 0.465$ and $\nu_0 < 0.1019$, or 16 $\nu_0$ =1.63 which means, that there is at most one solution to $F_{ex}$.

If we add the clause ( + 0 + 0) to $F_{ex}$ , the new instance $F_{ex} \wedge (a_1 \vee a_3) \in$ UNSAT, i.e. it will have no solution at all. But in this case we get $M$=2 ($a$=29/14) and $\nu_\leq = \nu_1 + \nu_2 = 13/16$ which leads to an even worse $x_{app}$=0,065183 and two solutions as the best guess from the inequality. Obviously, the choice of the approximation is important.

## V. Optimized cutoff.

Slightly better results are obtained with an optimized cutoff $a$.

Consider the r.h.s. of (III. 2) as a function of $a$:

(V. 1)   $f(a) = 1 - \dfrac{1}{a^2}a^2\Delta_\leq(a) - \dfrac{E(u)^2}{E(u^2) + a^2\Delta_\leq(a)}$ .

If we assume $M \leq a < M+1$ with $M \in \mathbb{N}$ then it is clear that $a^2 \Delta_{\leq}(a)$ is a continuous function of $a$ and so is $f(a)$. Furthermore $f$ is differentiable in the interval $M \leq a < M+1$, with the result

(V. 2)    $\dfrac{a}{2} \dfrac{df}{da} = \Delta_{\leq} - \nu_{\leq}(1 - \dfrac{a^2 E(u)^2}{(E(u^2) + a^2 \Delta_{\leq})^2})$ .

General statements about the slope of $f$ are possible and discussed in Appendix A.

In particular, for $M = 1$ one can show (see App. A) that the optimal bound is given by

(V. 3)    $\nu_0 \leq f_{min} = 1 - \nu_1 - \dfrac{(E(u) - \nu_1)^2}{E(u^2) - \nu_1}$ .

The r.h.s. is a monotonically decreasing function of $\nu_1$ starting from $1 - 1/\beta$ at $= \nu_1 = 0$ and reaching zero at $\nu_1 = \dfrac{\sigma_u^2}{\sigma_u^2 + (E(u)-1)^2}$ which is exactly the Cantelli-limit (II.6).

We check (V. 3) on the instance $F_{ex}$, as before. We get $f_{min} = 3/38 = 0.078..$ And thus $16\nu_0 < 1.263$ which predicts one solution correctly.

Now also the threshold for satisfiability is found correctly. If we add the clause $(+ 0 + 0)$ to $F_{ex}$, thus generating an UNSAT instance $F_{ex} \wedge (a_1 \vee a_3)$, we get $f_{min} = 3/136 = 0.022..$ and therefore $\nu_0 < 0.352../16$ for this instance, implying that there is no solution indeed. In contrast the basic bound (II.3) yields $16\nu_0 < 2.48..$ , wrongly allowing for 2 solutions.

**VI Large number of clauses.**

In appendix C we derive estimates for generic instances with many clauses, $m \gg 1$. We exploit the fact that both $\sigma_u^2$ and $E(u)$ are $O(n)$ quantities. Both can be calculated in p.t.. The simple bound (II. 3) then leads to

(VI. 1)        $\nu_0 \leq \dfrac{\sigma_u^2}{E(u)^2} + O(1/n^2)$

for large instances.

The improvements over the basic bound can lead to slightly better results, but do not remove the general problem that $\nu_0$ in general is of order $1/n$. For the optimized bound given in equation (V. 3), e.g., one gets

(VI. 2)        $\nu_0 \leq \dfrac{\sigma_u^2}{E(u)^2} - \nu_1 + O(1/n^2)$

and for the approximation of (IV. 3) the large $m$ - limit for $\nu_0$ reads

(VI. 3)        $\nu_0 \leq \dfrac{\sigma_u^2}{E(u)^2} - \dfrac{4\alpha}{E(u)^2} \nu_{\leq} + O(1/n^2)$

where $\alpha$ is a cutoff-dependent factor between 0 and 1 (namely $(a\text{-}M)^2$, see IV. 3)), and $\nu_{\leq}$ is defined in equ. (IV. 3).

Even if (VI. 2) and (VI. 3) yield results of the order of $1/2^n$ for $\nu_0$ in special cases, they do not allow for a determination of satisfiability in p.t. because of the presence of the probabilities $\nu_{\leq}$ and $\nu_0$.

## VII. Conclusion.

Equations (II. 3), (V. 3) and (VI. 1-3) may be considered as the main results of this paper. They represent bounds on the number of solutions to a given CNF instance. The bounds are expressed via the variance and the expectation value of $u(x)$, where $u(x)$ is the number of unsatisfied clauses for an assignment $x$. These quantities can be calculated in p.t.. However, only inequalities (II. 3) and (VI. 1) allow for a purely p.t. calculation, the others contain further characteristics of the distribution, namely the probabilities $\nu_1$ and $\nu_{\leq}$, which in general are not p.t. calculable.

The bounds on the number of solutions presented here are not apt to handle either SAT or #SAT problems in general. The reason is that the random variable $u$ is not able to resolve $O(2^{-n})$-distances in solution space. In particular for a determination of satisfiability in the sense of (II. 2) the formulas are not decisive in general, because – as can be seen most clearly in the large $m$ limit – they operate on a coarse-grained scale $O(1/n)$ instead of $O(1/2^n)$. For "small" instances or instances with $O(2^n)$ many solutions they might yield sensible results, however. We have checked this point for randomly generated instances of medium size with up to 60 clauses with inconclusive results.

## Appendix A.

(V. 2) can also be written

(A1) $\qquad \dfrac{a}{2}\dfrac{df}{da} = -\dfrac{1}{a^2}\sum_{k=1}^{M}\nu_k (k^2 - \dfrac{a^4 E(u)^2}{(E(u^2)+a^2\Delta_{\leq})^2})$ .

Note first that the factor in brackets is positive or zero for all $a^2 \leq E(u^2)/\mathrm{E}(u)$ . The same is true, independent of the value of a, if $E(u^2) < E(u)$ . Proof: because of (III. 4) one has

$\dfrac{a^2 E(u)}{E(u^2)+a^2\Delta_{\leq}} \leq \beta E(u) = E(u^2)/E(u) < 1$ . Thus the term in brackets in (A1) is positive for all $k$ if

$E(u^2) < E(u)$ and arbitrary $a$ or for $a^2 \leq E(u^2)/\mathrm{E}(u)$ . Thus (A1) does not exclude $\dfrac{df}{da} > 0$ for some

value of $a$ larger than $\sqrt{E(u^2)/\mathrm{E}(u)} > 1$. In this case $f$ would have a local minimum which improves the bound.

To see that this can indeed be the case, we choose $1 < a < 2$ , i.e. $M$=1. Then $\Delta_{\leq} = \nu_1(1-1/a^2)$

and (A1) becomes zero at $a_{min}^2 = \dfrac{E(u^2) - v_1}{E(u) - v_1}$ . According to the second derivative at this point,

$\dfrac{d^2 f}{da^2} = \dfrac{8 v_1}{a_{min}^4}(1 - \dfrac{v_1}{E(u)}) > 0$ , this is a local minimum indeed . Then the minimum value of $f$ is given by

(A2)     $f_{min} = 1 - v_1 - \dfrac{(E(u) - v_1)^2}{E(u^2) - v_1}$     .

To see the improvement over the basic bound (II.3) , $v_0 \leq 1 - 1/\beta$, one writes $f_{min}$ in the form

(A3)     $f_{min} = 1 - \dfrac{1}{\beta}\left(1 + v_1 \dfrac{\left(E(u^2)/E(u) - 1\right)^2}{E(u^2) - v_1}\right)$

(A2) is also an improvement over

$$v_0 \leq 1 - v_1 - \dfrac{(E(u) - v_1)^2}{E(u^2)}$$

which follows directly from the Cauchy-Schwartz inequality (see equ. (II. 4)) by choosing $A = u$ and $B = 1$ for $u > 1$ and zero otherwise.

## Appendix B.

To show that the cutoff formulation gives a better estimate than the simple CS- estimate we observe that     $\Delta_{\leq} = E(1 - \dfrac{u^2}{a^2}) - v_0 + 2^{-n}\sum_{S_>}(\dfrac{u^2}{a^2} - 1) = 1 - \beta \mu^2 - v_0 + |\Delta_>|$ .

If one drops the last positive term and uses (II. 3) as an estimate for $v_0$ one gets

$$\Delta_{\leq} > \dfrac{1}{\beta} - \beta \mu^2$$

for the actual $\Delta_{\leq}$ for any choice of $a$. Thus the last term in equ. (III. 2) gives a negative contribution to $1 - 1/\beta$ independent of $a$.

## Appendix C.

Any given SAT instance can be transformed to CNF $\cap$ 3-SAT $\cap$ READ-3 in p.t. After removing pure literals and unit clauses, which are trivial complications, one is left with a CNF formula, whose clauses have either 2 or 3 literals, and in which each atom appears either 2 or 3 times. Furthermore one can assume that the appearance of each atom $a$ is either of the form $(a \vee X) \wedge (a \vee Y) \wedge (\bar{a} \vee Z)$ or $(a \vee X) \wedge (\bar{a} \vee Y)$ ; otherwise it had been removed as a pure literal. An appearance of the form $(a \vee X) \wedge (\bar{a} \vee Y) \wedge (\bar{a} \vee Z)$ can be changed by flipping $a$ , a transformation which does not alter the number of solutions. Let us call such formulae "slim-SAT" formulae. Obviously slim-SAT $\in$ NP. To discuss the large-$m$ behaviour of a SAT instance it is therefore sufficient to consider slim-SAT formulae without loss of generality. Slim-SAT formulae are (not completely) characterized by a couple of numbers: they have $m_3$ clauses with 3 literals, $m_2$ clauses with 2 literals, $n_3$ atoms which appear 3 times, $n_2$ atoms which appear 2 times. If we denote the total number of literals in the formula by $N$ the following relations hold

$$m_2 = 3m - N \;\; ; \;\; m_3 = N - 2m \;\; ; \;\; n_2 = 3n - N \;\; ; \;\; n_3 = N - 2n \;\; .$$

Furthermore $n$ equals the number of negative literals, and $N$-$n$ the number of positive literals.

From these relations it is not difficult to derive the inequality

$$\frac{2}{3}n \le m \le \frac{3}{2}n \;\; .$$

Therefore the size of a formula may be characterized by $m$ as well as by $n$, and O($n$)=O($m$).

Now consider the terms in (I. 4a and b). Obviously for slim-SAT $C = E(u) = m_3/8 + m_2/4 = $O($m$)=O($n$). We can also draw conclusions on the number of $\lambda$-, $\mu$-, and $\nu$-terms in (I. 4a) and thus in (II. 9) from the structure of slim-SAT formulae. Obviously the number of different $\lambda$-terms is $n$; furthermore the number of $\mu$-terms cannot be larger than $3m_3 + m_2$ and therefore must be smaller than $3n$, and the number of different $\nu$-terms is limited by $m_3$, i.e. also O($n$). Therefore $\dfrac{\sigma_u^{\,2}}{E(u)^2}$ is an O($1/n$)-quantity which can be used as an expansion parameter and justifies the approximations leading to (VI.1-3).

**References and remarks.**

 

$$E(AB)^2 \le E(A^2)E(B^2) - (\sigma_A E(B) - \sigma_B E(A))^2 \;\; .$$

Completing the square on the r.h.s. yields the identical

$$\left| E(AB) \right| \leq \sigma_A \sigma_B + E(A)E(B)$$

Applied to our problem with functions *A* und *B* as defined following equation (II. 4) one gets exactly the basic bound (II. 3), which follows immediately from CS. The reason why Walkers formula does not lead to an improvement of the CS-inequality in our case is that it appears to be a straightforward <u>consequence of CS</u> applied to variables *A-E(A)* and *B-E(B)*.